\documentclass[aps,prd,floatfix,superscriptaddress,twocolumn,preprintnumbers,floatfix]{revtex4} 
\usepackage{amssymb}
\usepackage{amsmath}
\usepackage{amsfonts}
\usepackage{epsfig}             
\usepackage{graphicx}
\usepackage{stackrel}
\usepackage{tabularx}
\usepackage{color}
\usepackage{dsfont}
\usepackage[T1]{fontenc}
 \usepackage{dblfloatfix}
 \usepackage{float}

\usepackage{lipsum}

\definecolor{darkerblue}{rgb}{0.0,0.0,0.5}
\newcommand{\seq}{\begin{subequations}}
\newcommand{\sen}{\end{subequations}}

\newcommand{\eq}{\begin{eqnarray}}
\newcommand{\en}{\end{eqnarray}}

\def\nn{\nonumber}

\begin{document}
	
	\title{Probing invisible vector meson decay mode\\
		 with  hadronic beam in the NA64 experiment at SPS/CERN} 

	\author{Alexey~S.~Zhevlakov \footnote{{\bf e-mail}: zhevlakov@theor.jinr.ru}} 
\affiliation{Bogoliubov Laboratory of Theoretical Physics, JINR, 141980 Dubna, Russia} 
\affiliation{Matrosov Institute for System Dynamics and 
	Control Theory SB RAS, \\  Lermontov str., 134, 664033, Irkutsk, Russia } 

\author{Dmitry V.~Kirpichnikov \footnote{{\bf e-mail}: kirpich@ms2.inr.ac.ru}}
\affiliation{Institute for Nuclear Research of the Russian Academy 
	of Sciences, 117312 Moscow, Russia} 
	
\author{ \\ Sergei N.~Gninenko \footnote{{\bf e-mail}: Sergei.Gninenko@cern.ch}}
\affiliation{Institute for Nuclear Research of the Russian Academy 
	of Sciences, 117312 Moscow, Russia} 

\author{ Sergey Kuleshov \footnote{{\bf e-mail}: sergey.kuleshov@unab.cl}} 
\affiliation{Millennium Institute for Subatomic Physics at
	the High-Energy Frontier (SAPHIR) of ANID, \\
	Fern\'andez Concha 700, Santiago, Chile}
\affiliation{Center for Theoretical and Experimental Particle Physics, Facultad de Ciencias Exactas, Universidad Andres Bello, Fernandez Concha 700, Santiago, Chile}
		
	\author{Valery~E.~Lyubovitskij \footnote{{\bf e-mail}: valeri.lyubovitskij@uni-tuebingen.de }} 
	\affiliation{Institut f\"ur Theoretische Physik, Universit\"at T\"ubingen, \\
		Kepler Center for Astro and Particle Physics, \\ 
		Auf der Morgenstelle 14, D-72076 T\"ubingen, Germany} 
	\affiliation{Departamento de F\'\i sica y Centro Cient\'\i fico
		Tecnol\'ogico de Valpara\'\i so-CCTVal, \\ 
		Universidad T\'ecnica Federico Santa Mar\'\i a, Casilla 110-V, Valpara\'\i so, Chile}
	\affiliation{Millennium Institute for Subatomic Physics at
		the High-Energy Frontier (SAPHIR) of ANID, \\
		Fern\'andez Concha 700, Santiago, Chile}
	
	\date{\today}
	
	\begin{abstract}

We test a novel idea of using  a  $\pi^-$ beam in the fixed-target experiments 
to search for  New Physics in the events with missing energy. 
Bounds for invisible vector $\rho$ meson decay were derived, analyzed, 
and compared with the current limits on searching Dark Matter in the  
accelerator based experiments.
We demonstrate that the new approach can be effective tool to probe  
sub-GeV Dark Matter parameter space. 

	\end{abstract}
		
	\maketitle
	
	\section{Introduction}
 \label{Intro}

Searching for Dark Matter (DM) is well-motivated challenge 
in particle physics that stimulates experimental and 
theoretical efforts for decades. 
Study of DM phenomenology gives a unique opportunity 
to explain many observations in astrophysics and cosmology. 
In a wide range of possible DM candidates, 
we can mention light DM in the sub-GeV mass region 
which could potentially explain several 
observed anomalies~\cite{Boehm:2003hm,Gunion:2005rw} and could be a candidate for thermal relic dark sector. 
An idea of dark portals between the hidden and ordinary matter,  
described by the  Standard model (SM), typically implies 
light sub-GeV intermediate states.  
In particular, there are several hidden sector scenarios 
that have been widely discussed in literature: 
the Higgs portal~\cite{Arcadi:2019lka,Davoudiasl:2021mjy}, 
the tensor portal~\cite{Voronchikhin:2023znz,Voronchikhin:2022rwc,Kang:2020huh}, 
the dark photon portal~\cite{Fortuna:2020wwx,Buras:2021btx,Kachanovich:2021eqa}, 
sterile neutrino portal~\cite{Escudero:2016tzx}, 
axion or axion-like (ALPs)  portals~\cite{Nomura:2008ru,Zhevlakov:2022vio}. 
In addition, we note that such models are considered typically
in the framework of lepton-specific~\cite{Sieber:2023nkq} or hadron-specific 
cases~\cite{Zhevlakov:2022vio}.

The scenarios with dark portal states predict missing energy events in reactions 
with leptons~\cite{Buras:2021btx,Kachanovich:2021eqa,Radics:2023tkn} and 
hadrons~\cite{Dreiner:2009er,Badin:2010uh,Fayet:2006sp,Bauer:2020jbp,Goudzovski:2022vbt}, including lepton flavor violation effect~\cite{Zhevlakov:2023jzt,Bauer:2021mvw}.  
The invisible decays play an important role in testing SM and searching for DM particles. 
Experimental studies of invisible hadronic decays were performed by several collaborations. 
In particular, BES III Collaboration~\cite{BESIII:2012nen,BESIII:2018bec} set the constraints on the invisible branching fraction of the $\eta$, $\eta'$, $\omega$, and $\phi$ mesons.  
BABAR Collaboration~\cite{BaBar:2009gco,BaBar:2013npw} has been studied the invisible decay modes of heavy quarkonia. NA62 Collaboration~\cite{NA62:2020pwi} established the limits on invisible decays of $\pi^0$. 
Existed  limits on DM from invisible decays of the vector DM mediator~\cite{Workman:2022ynf} derived from analysis of data collected in the $e^+e^-$ colliders and accelerator based experiment NA64~\cite{NA64:2023wbi,Gninenko:2023pkv}.
Experiments which aimed for direct DM detection and probing meson decay into invisible mode may provide an important signatures of sub-GeV DM~\cite{Arefyeva:2022eba}. 
Invisible meson decays can be limited by using missing energy/momentum 
techniques~\cite{Gninenko:2015mea,Gninenko:2016rjm}. 
In the framework of missing energy concept bounds to invisible decay into DM were obtained 
in~\cite{Schuster:2021mlr} for such experiments as NA64 and LDMX where vector mesons are created by interaction of radiated photons from electron beams in the calorimeter.
Many existed and future experiments for searching of DM based on a use of 
missing energy/momenta techniques are concentrated on setups with lepton beams colliding 
fixed atomic targets. Here, the main aim is to search for missing energy/momenta signal 
events which can be interpreted as potential signatures of the produced DM. 

In the present paper we extend the analysis of invisible meson decays 
to DM by using missing energy conception which was considered previously 
in Ref.~\cite{Schuster:2021mlr}. Our main objective is to test the potential 
of missing energy techniques for invisible meson decay for hadronic beams. 
NA64 Collaboration has been started to exploit this concept in the experiments with 
hadronic beams to search for signatures of dark matter 
production~\cite{Crivelli:2023pxa,Antel:2023hkf}. 
For the first time the experiment will use the beam $\pi^-$ mesons scattered 
at the active target. 
During last two years (runs in 2022 and 2023 years with a few days of data collection), 
NA64 Collaboration accumulated about $3 \times 10^{9}$ pions on target in order 
to understand potential of the NA64 detector by using pion beam and 
missing energy technique.
Another aim of our paper is to estimate a sensitivity of hadronic-pion beam 
to search for DM implementing missing energy techniques. 
In particular, we will make an estimate of observable in invisible meson decays. 
In our analysis we rely on preliminary analysis of accumulated number of pions 
on target from NA64 technical run ($3\times 10^{9}$) and make predictions for 
the  projected statistics in the range between  $5 \times 10^{12}$ and $ 10^{14}$  
pions on target. 

The paper is organized as follows. In Sec.~\ref{Framework} we describe missing energy conception to analyze invisible vector meson decay mode for hadronic case beam of 
the NA64 experiment and estimate yield of vector meson in experimental facility 
which can be used for analysis. In Sec.~\ref{VMP} we calculate the cross section of 
neutral $\rho^0$ vector meson production in the $\pi^-$ scattering at nuclear target. 
The discussion about invisible meson decay mode to DM fermions and implementation to DM parameter space is presented in Sec.~\ref{bounds_sec}. 
Finally, in Sec.~\ref{Conclusion} we present our conclusions.
	
\section{Framework}
\label{Framework}

Missing energy conception, proposed in~\cite{Gninenko:2013rka},  
is pretty well realized and work consistently at fixed target experiments 
dealing with lepton (electron and muon) beams, such as the NA64$_e$ and 
NA64$_\mu$~\cite{Gninenko:2014pea} experiments. In future, it is planed to 
run several new experiments, e.~g.~, 
M$^3$~\cite{Kahn:2018cqs,Capdevilla:2021kcf} and 
LDMX~\cite{Berlin:2018bsc,LDMX:2018cma,Ankowski:2019mfd,Schuster:2021mlr,Akesson:2022vza}. 

For 90\% confidence level (C.L.) limit on the invisible branching ratio of 
produced meson, $V$, in experiment where we assume zero observed signal events and background free case, that implies $\mathrm{Br}(V\to \text{inv.})\leq 2.3/N_V$, where $N_V$ is a number of the produced vector mesons. We 
consider an modified experimental setup of the NA64 to estimate the 
$\rho^0$ meson production in the reaction of the $\pi^-$ beam scattered at the active iron target. 
About $3\times 10^{9}$ pions on target $(\pi \mbox{OT})$ in the NA64 experiment were accumulated in a short period  of technical data taking at SPS/CERN. 
For the typical projected statistics of NA64 we will use both number 
of $5 \times 10^{12}$ and $10^{14}$ pions on the target. 

\begin{figure}[b]
	\includegraphics[width=0.33\textwidth, trim={2cm 22cm 10.4cm 1.8cm},clip]{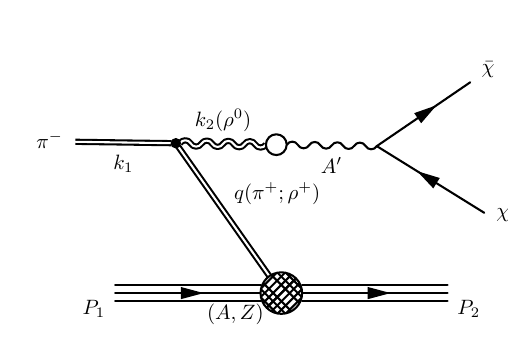}
	\caption{Feynman diagram describing the cascade process of the $\rho^0$ vector meson production due to the $\pi^-$ scattering at nuclear target followed by the transition 
    to the dark photon and DM fermions.}
	\label{DiagRhoProd_dominant}
\end{figure}

The cross section for the $\rho^0$ meson production in the reaction 
$\pi^- + (Z,A) \to \rho^0 +(Z-1,A)$ is given by the formula: 
\eq
&&\sigma(\pi^- + (Z,A) \to \rho^0 +(Z-1,A))\nn\\&& \qquad= Z \sigma(\pi^- + p \to \rho^0 +n) 
\,.
\en 
Here we assume that the main channel of the $\rho^0$ production is due to positive pion 
captured by the nuclear target (see Fig.\ref{DiagRhoProd_dominant}). 
This process is the dominant one and occurs due to the $\pi^+$ exchange in the $t$-channel. 
Second possible mechanism for the $\rho^0$ production could occur due 
annihilation of $\pi^-$ from the beam with $\rho^+$ or two-pion pair 
$(\pi^+ \pi^0)$ radiated from the target.  We will show below  that 
the latter mechanism is strongly suppressed in comparison with 
the leading signature of the $\pi^+  \pi^-$ annihilation. 

To constrain the parameters of dark photon coupled 
with vector mesons~\cite{Schuster:2021mlr}, 
we need to estimate the invisible branching ratio  
$\mathrm{Br}(V\to \text{inv.})$ for vector meson production. 
Here we will focus on dark sector with pseudo-Dirac DM fermions, 
which couple to the U(1)$_D$ vector mediator (dark photon). 
Such a coupling is described by Lagrangian 
\eq
\mathcal{L}\supset \epsilon e A'_\mu J^{\mu} 
+ g_D A'_\mu \bar{\chi}\gamma^\mu \chi \,, 
\en
where $\epsilon$ is the kinetic mixing parameter~\cite{Holdom:1985ag},
$g_D$ is the coupling of dark photon 
with dark fermions,
$e$ is the electric charge, and $J_\mu$ the electromagnetic current 
composed of the SM fermions. In the following we use the notation
of the effective dark coupling constant $\alpha_D =g_D^2/4\pi$. 
The coupling of vector meson $\rho^0$ meson with dark photon is defined 
by analogy with QED photon but it has an extra factor $\epsilon$ 
(kinetic mixing coupling). 
The width of the decay of vector meson into the dark fermion pair 
$V\to \bar{\chi}\chi$ is given by 
\eq
\Gamma_{V\to \bar{\chi}\chi }  = \frac{g_D^2 (\epsilon e)^2}{12\pi} g_V^2 \frac{(m_V^2+2m_\chi^2)\sqrt{m_V^2-4m_\chi^2}}{(m^2_{A'}-m_V^2)^2+\Gamma^2_{A'\to \bar{\chi}\chi }m^2_{A'}} \,, 
\en
where $g_V$ is the vector meson meson coupling with current, 
$m_{A'}$ and $m_\chi$ are the masses of intermediate dark 
photon and pseudo-Dirac DM fermion, respectively, 
$m_V$ is the mass of vector meson. Here we use the Breit-Wigner 
propagator for the dark photon $A'$ assuming  that its total width 
is dominated by the $A' \to \bar{\chi}\chi$ mode. 
Decay width $\Gamma_{A'\to \bar{\chi}\chi}$ is 
\eq 
\Gamma_{A'\to \bar{\chi}\chi } = \frac{g_D^2}{12 \pi} 
\, m_{A'} \, 
(1 + 2 y_\chi^2) \, (1 - 4 y_\chi^2)^{1/2} \,,  
\en 
where  $y_\chi = m_\chi/m_{A'}$.

Number of vector mesons produced by $\pi^-$ beam scattering 
at fixed target is
\begin{equation}
N_{\rho} \simeq \pi\mbox{OT}\cdot \frac{\rho_{T} N_A}{A} L_T \int\limits^{\theta_{max}}_{0}
d\theta \frac{d \sigma_{2\to2}}{d\theta}\,,
\label{Nrho}
\end{equation}
where $A$ is the atomic weight number, $N_A$ is the Avogadro's number, $\pi\mbox{OT}$ is the number of negative charged pions accumulated on target, $\rho_{T}$ is the target density, $L_T$ is the effective thickness of the
target which in conservative scenario is assumed to be equal to
effective pion interaction length in the target~\cite{Workman:2022ynf},
$d\sigma_{2\to2}/d\theta$ is the differential cross section of
the $\rho^0$ meson production process, $\theta$ is a angle between $\pi^-$ beam line
and the momentum of the produced $\rho^0$ meson.  

\begin{figure}[b]
	\includegraphics[width=0.49\textwidth,clip]{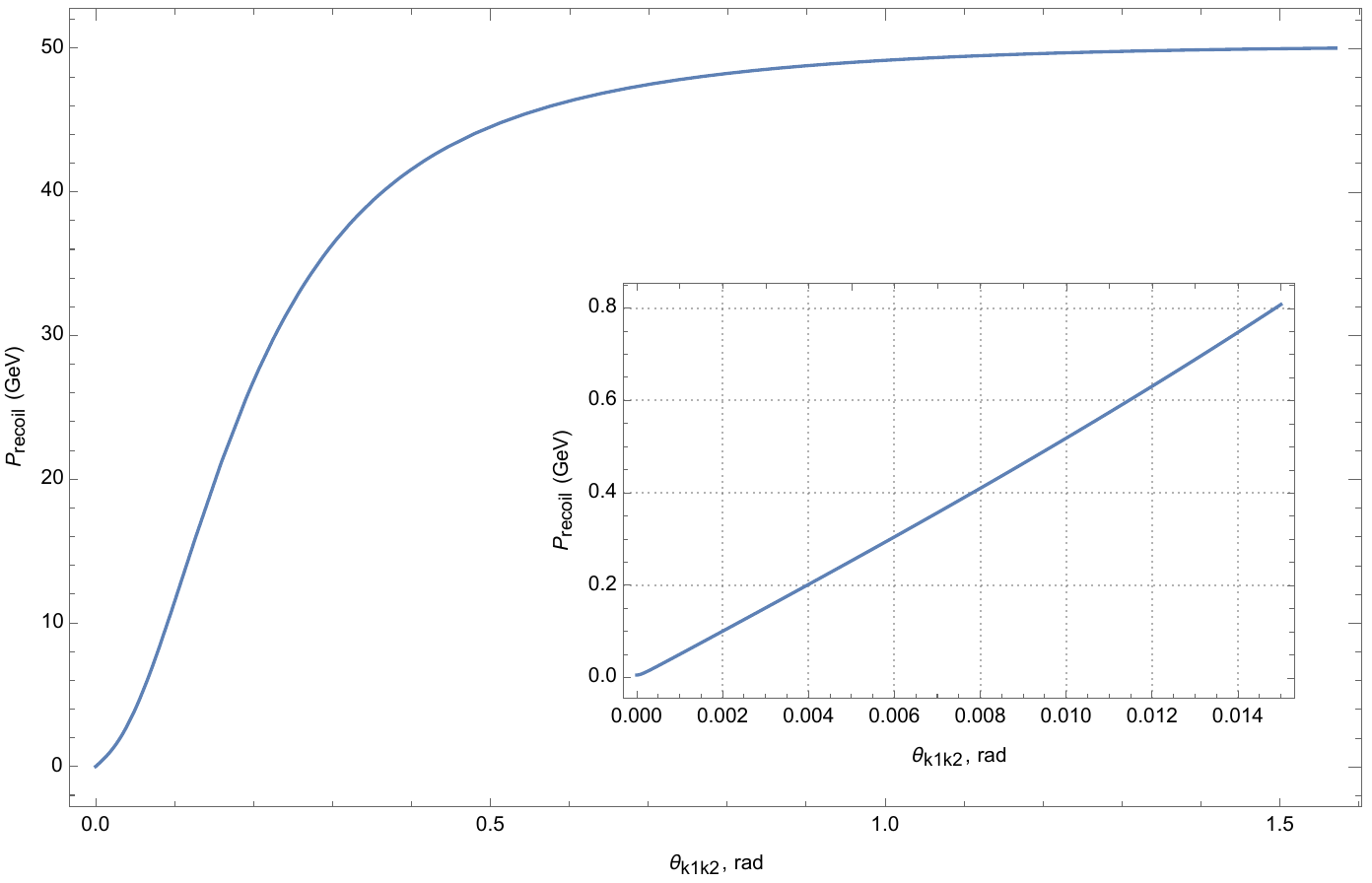}
	\caption{Recoil momenta to nucleon at beam energy $E_{\pi^-} = 50$ GeV 
    with creation $\rho$ meson in final state. The small figure shows 
    the area where recoil momenta to nucleons less than 1 GeV and can be used for calculation and analysis of missing energy signals.}
	\label{Recoil_E}
\end{figure}

The cross section of $\rho^0$ meson production plays an important role in calculation of vector mesons flow and crucially depends on the angle $\theta_{max}$.  
Maximum of the scattering angle $\theta_{max}$ is defined by experimental cuts of 
registration of signal which can be interpreted as missing energy by transition 
to DM. For the NA64 experiment it is important to maintain a negligible 
background in a calorimeter system. The latter can be provided by the adjusting the recoil energy from final neutrons or pieces of the disintegration of the atomic 
nucleus that implies the energy deposition inside the hadronic calorimeter. To take 
into account the detector response one may fix the $\theta_{max}$ from minimal possible 
background energy emission from recoil energy of the final neutron. Dependence of the recoil momentum of the final neutron on the scattering angle is shown in  Fig.~\ref{Recoil_E}. We set an optimistic upper limit 
on the recoil momentum of the neutron to be $0.8$ GeV. 
This limit corresponds to $\theta_{max}\sim 0.014$~rad for the 50 GeV pion beam. 
For the 100 GeV pion beam, we need to use $\theta_{max}\sim 0.008$~rad. 
By using those cuts on the scattering angle we obtain the optimal
missing energy cut and relatively small background which can be suppressed experimentally. This small area of angle which can be used for analysis of missing energy is connected to the kinematic of scattering of massive particles in initial 
and final states. In addition, it is worth noticing a difference between our analysis and the study presented in Ref.~\cite{Schuster:2021mlr} for electron beam experiments that provides the bounds on pseudo-Dirac DM from invisible vector meson decays. 
In Ref.~\cite{Schuster:2021mlr} vector mesons are produced inside the 
calorimeter by interaction of a bremsstrahlung photon with matter of the calorimeter. We note that a small typical angle of outgoing $\rho^0$ meson decreases sensitivity of hadron missing energy experiment. But for a more accurate 
analysis one needs to carry out a proper Monte Carlo simulation of this process in detector that includes also the background from recoil neutrons. 
For the larger angle cut  one  needs to take into account nuclear function and all possible transitions of the nucleus during the transfer of 
energy to the nuclear shell. We keep the analysis a full possible picture of 
hadronic showers in the detector for future  study. 

We need to point out that the potential background for the invisible decays of mesons  can arise from the decay of neutral mesons into neutrino-antineutrino pair. 
These decays are strongly suppressed from SM and the regarding decay widths are estimated to be at the level of $\Gamma(M^0\to\nu\bar{\nu})
\lesssim 10^{-16}$~\cite{Gninenko:2014sxa}. Remarkably, that for $\rho^0$ meson the typical bound can be set as follows  
$\Gamma(\rho^0\to\nu\bar{\nu})\lesssim 4.2 \times 10^{-13}$ \cite{Gao:2018seg}. 
However, an experimental signatures for such decays have not been observed yet.
An existence of any experimental evidence for invisible meson decay can be considered as a potential signal of New Physics.   

The yield of neutral vector $\rho^0$ mesons at the NA64 experiment with $\pi^-$ beam 
is shown in Table.~\ref{ParamTable} for $E_{beam} = 50$ and 100 GeV. 
Besides, in Table.~\ref{ParamTable} we also show the typical fraction 
$f(\theta_{max})$ of $\rho^0$ mesons that can be produced within two benchmark angle ranges $\theta \lesssim \theta_{max}=0.014$ rad 
and $\theta \lesssim \theta_{max}=0.022$ rad. It correspond to the typical neutron recoil momenta at the level of 0.8 GeV and 1.2 GeV, respectively. 
The target of the experiment is a hadronic calorimeter which 
represents four or three modules in 48 layers (2.5 mm of iron plates and 4 mm of scintillator). The possible signature of the neutron recoil momentum of 1.2 GeV can be deposited in the hadronic calorimeter at the level of $\simeq 1$ GeV, it implies  $\simeq 0.2$ GeV is transferred to the nucleus as a typical energy of the 
nucleus excitation. It is important to note, that 
small recoil energy of neutron can be achieved by decreasing the energy of the pion beam. This scenario for 20 GeV of pion beam 
is shown in Table.~\ref{ParamTable} for the neutron recoil momentum of 1 GeV. 

\onecolumngrid
\begin{center}
	\begin{table*}[t]
		\centering
		\caption{Parameters of the fixed-target experiments NA64$_h$ for the iron  target ($A=56, Z=26$, $\rho=7.874$ ( g cm$^{-3}$), and interaction length of pion in Fe $L_T=20.41$ cm): $E_{beam}$ is the beam energy of pions, $\sigma_{tot}$ is a total cross section, $f(\theta_{max})$ is  fraction of $\rho^0$ mesons which are produced within small angle range, $\theta \lesssim \theta_{max}$, $\pi$OT is a typical number of pions on accumulated on target, $N_{\rho}$ is  yield of neutral $\rho^0$ vector mesons. 
		}
		\begin{tabular}[b]{rcccccccc}
			\hline
			\hline
			 & $E_{beam}$ (GeV)&$\sigma_{tot} ({\mu}b) $& $f({\theta_{max}})$& $\pi$OT & \quad $ \theta_{max}$&\quad$N_\rho$\\
			\hline
			\\
			NA64$_h$: \quad &50 & 0.113 & $3.1 \times 10^{-6}$ & 3 $\times $ $ 10^{9}$& 0.014&\quad 1.8 $\times $ $  10^{3}$ &\\
			NA64$_h$: \quad&100 & 0.117 & $4.7 \times 10^{-7}$ & 3 $\times $ $10^{9}$&0.008 &\quad 0.29 $\times $ $  10^{3}$&\\
			NA64$_h$: \quad&50&0.113& $3.1 \times 10^{-6}$&5 $\times$ $10^{12}$ &0.014 &\quad 3.1  $\times$  $  10^{6}$&\\
			NA64$_h$: \quad&100&0.117&$4.7 \times 10^{-7}$&5 $\times$ $10^{12}$  &0.008 &\quad 0.48  $\times $ $  10^{6}$&\\
			\hline
            NA64$_h$: \quad &50 & 0.113 &$ 7 \times 10^{-6}$ & 3 $\times $ $ 10^{9}$& 0.022&\quad 4.1 $\times $ $ 10^{3}$ &\\ 
            NA64$_h$: \quad &50 & 0.113 &$ 7 \times 10^{-6}$ & 5 $\times  $ $10^{12}$& 0.022&\quad 6.9 $\times $ $ 10^{6}$ &\\
            \hline
            NA64$_h$: \quad &20 & 0.104 &$ 6.7 \times 10^{-4}$ & 3 $\times $ $ 10^{9}$& 0.05&\quad 5.2 $\times $ $ 10^{4}$ &\\ 
            NA64$_h$: \quad &20 & 0.104 &$ 6.7 \times 10^{-4}$ & 5 $\times $ $ 10^{12}$& 0.05&\quad 8.7 $\times $ $ 10^{7}$ &\\ 
			\hline
                \hline
			\label{ParamTable}
		\end{tabular}
	\end{table*}%
\end{center}
\twocolumngrid

In our calculations we use formulas for the cross-section of the $\rho^0$ meson production which is calculated in the next section. Estimate 
of the $\rho^0$ production is obtained for current and ultimate statistics 
and for two possible values of the pion energy in a beam in a narrow angle of meson production.  
 
\section{vector meson production} 
\label{VMP}

In this section we briefly discuss formalism and obtain an expression for  differential cross section of $\rho^0$ vector meson production in the  $\pi^- + p \to \rho^0 +n$ reaction.  Our formalism is based on Lagrangians that includes nucleons $N=(p,n)$, pseudoscalar mesons $\pi^\pm$, vector $\rho^\mu$  mesons and $A_\mu$ photon.
\eq
{\cal L}_{\pi NN} &=& g_{\pi NN} \bar{N} i\gamma_5 \vec{\pi\,} \vec{\tau\,} N\,,\\
{\cal L}_{\rho \pi \pi} &=& i g_{\rho \pi \pi} \rho^\mu(\partial_\mu \pi^\dag \pi-\pi^\dag \partial_\mu  \pi)\\
{\cal L}_{\rho NN} &=&g_{\rho NN} \bar{N} \gamma_\mu \rho^{\mu,a} \vec{\tau}_a N\,,\\
{\cal L}_{\pi \rho\rho} &=& g_{\pi \rho \rho} \pi F_{\rho}^{\mu\nu} \tilde{F}_{\rho}^{\alpha\beta} \,. 
\en	
Here $F_{\rho}^{\mu\nu}$ and $ \tilde{F}_{\rho}^{\alpha\beta}$ are the strength 
tensors and dual tensor of vector meson field, respectively, $\gamma_\mu$ and $\gamma_5$ are Dirac matrices. 
The couplings occurring in the above equation are
\eq
g_\rho&=&2F_\pi^2 g_{\rho\pi\pi}\,, \\ 
g_{\rho\pi\pi} &=& \frac{m^2_\rho}{2 F_\pi^2} \,, 
\en
where $g_{\rho NN}\sim g_{\rho\pi\pi}$, $g_{\pi\rho\rho}=g_\rho^2/4\pi F_\pi$, $g_{\pi NN} =g_A m_N/F_\pi$, $g_A \approx 1.275$ is the nucleon axial charge and $F_\pi\approx 92.4$ MeV is the pion decay constant~\cite{Fujiwara:1984mp,Workman:2022ynf,CurrentHPBook}. Lagrangian 
describing transition of neutral vector meson to dark photon 
\eq
{\cal L}_{\rho-A'} &=&  e  \epsilon g_{\rho} m_\rho^2  \rho_\mu A'^\mu,
\en
can be derived from Lagrangian defining the $\rho-\gamma$ 
coupling~\cite{Jegerlehner:2011ti,Kachanovich:2021eqa} 
using well-known shift of electromagnetic field $A^\mu \to A^\mu + \epsilon A'^\mu$. 
In this work we consider two different channels of the $2\to2$ processes with $\rho^0$ vector meson production. The dominant channel is induced by the exchange of 
the $\pi^+$ mesons radiated off target in the $t$-channel. The second channel 
occurs due the $t$-channel exchange of the $\rho^+$ or the $\pi^- \pi^0$ 
pair. Both channels are shown in Fig.~\ref{DiagRhoProd_dominant}.

One can consider $2\to2$ process in the approximation of a small scattering 
angle. In this case the Mandelstam variables are  
\eq
s+t+u&=&2m_N+m_\pi^2+m_\rho^2, \nn \\
t &=& m_\pi^2+m_\rho^2-2(k_1k_2)\nn\\
&\approx& m_\rho^2\frac{(x-1)}{x}+m_\pi^2(1-x)-\theta_{k_1k_2}^2xE_{k_1}^2 ,\nn\\
s&=&m_\pi^2+2m_NE_{k_1}+m_N^2,\\
u&=&m_\rho^2+m_N^2+2E_{k_1}m_N x, \nn
\en
and recoil energy for the final neutron is 
\eq
E_{p_2}=\frac{2m_N^2-t}{2m_N}
\en
where
$x=E_{k_2}/E_{k_1}$ is fraction of pion beam energy beam transferred  to the outgoing vector meson, that can be 
associated with the typical missing energy, $x \simeq  1$ for relatively small angle, $\theta \ll 1$, of the 
produced $\rho^0$, $E_{k_1}$ is energy of pion beam, $E_{k_2}$ is energy of  the dark photon, $m_N$, $m_\pi$ 
and $m_\rho$ are the nucleon, pion and $\rho$- meson masses, respectively. 

The differential cross-section for the $2\to2$ process is 
\eq
d\sigma_{2\to2} =  \frac{1}{2 (4 j)} \sum_{s(P_1)} \sum_{\lambda'} |M|^2 dF_2
\en
where $j=\sqrt{E_{k1}^2m_N^2-m_N^2 m_\pi^2}$ is invariant flow, $dF_2$ is 
the 
corresponding phase space factor: 
\eq
dF_2 = \frac{1}{16 \pi s} \lambda(s,m_\pi^2,m_N^2)^{\frac{1}{2}} \, 
d\cos\theta_{k_1k_2}
\en
\onecolumngrid
\onecolumngrid
\begin{figure*}[t]
	\includegraphics[width=0.497\textwidth,clip]{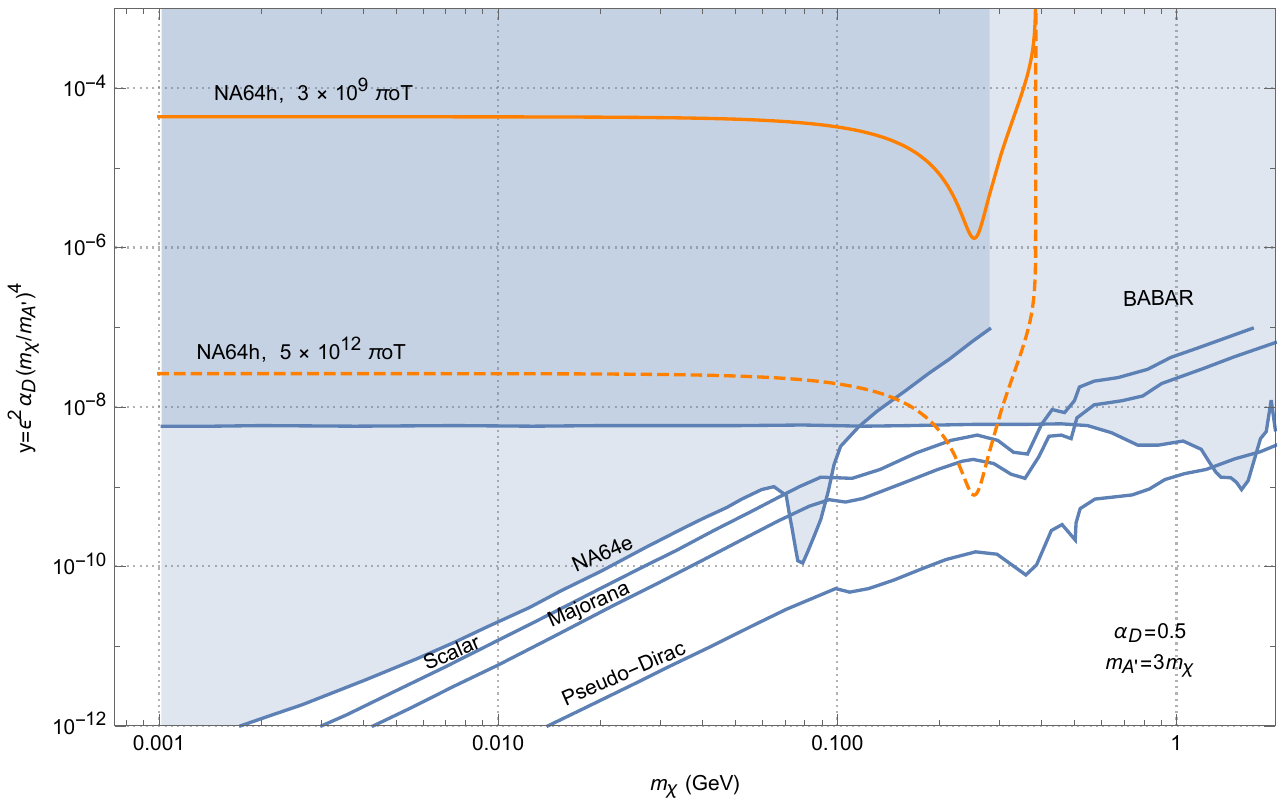}
	\includegraphics[width=0.497\textwidth,clip]{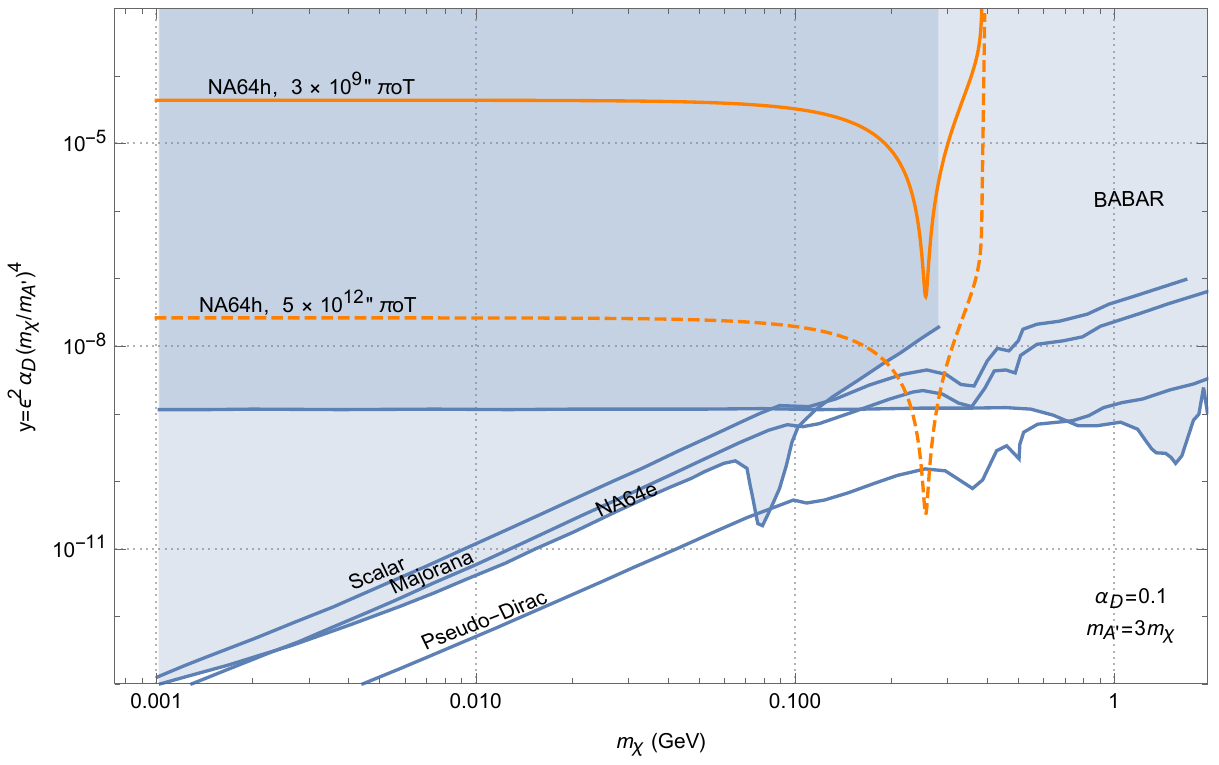}
	\caption{The constraints on parameters space of the dark photon mediator for pseudo-Dirac  DM. At both panels, we show existed limit from last data of NA64$_e$ experiment~\cite{NA64:2023wbi}, our estimates for the future 
NA64$_h$ experiment with a hadron beam \cite{Gninenko:2014sxa} and constraints from production of DM in $e^+e^-$ collision at BABAR \cite{BaBar:2013npw}. In  left panel we show  the limits for $\rho^0$ neutral vector meson invisible decay for current (few days of data taking) and  ultimate statistics of NA64$_h$ that implies  $\alpha_D=0.5$ and $m_{A'}=3m_\chi$. In right panel the same as in left panel but for  $\alpha_D=0.1$ and $m_{A'}=3m_\chi$.}
	\label{DiagDiffAlphaD}
\end{figure*}
\twocolumngrid
\twocolumngrid

where  $\lambda(x, y, z) = x^2 + y^2 + z^2 - 2xy - 2xz - 2yz$ 
is the K\"allen kinematical triangle function. 
The matrix element squared is provided below. We conservatively 
assume that  the maximum scattering angle of $\rho^0$ is determined by 
typical cuts of the NA64$_h$ experiment. 
The angle $\theta_{k_1k_2}$ is connected with  fraction $x$ from conservation laws of energy and momentum.  

\subsection{The dominant channel
\label{SecDominantCh}}	

For the dominant process with $\pi^+$ exchange, the  matrix element squared 
in the laboratory frame has the following form 
\eq
\frac{1}{2} \sum_{s(P_1)} \sum_{\lambda'} |M|^2 
&=&  \frac{g_{\pi NN}^2 g^2_{\rho \pi\pi}}{2(t-m_\pi^2)^2}
\frac{t}{2m_\rho^2}\Big[(m_\pi^2-t)^2\\
&&-m_\rho^2(2m_\pi^2-m_\rho^2-4m_NE_{k_1}(1-x))\Big] \,,\nn
\en
where the sum over the polarizations of massive vector boson is given by 
\eq
\sum_{\lambda_\rho}  \epsilon^\mu_{\rho}(\lambda_\rho)
\epsilon^\nu_{\rho}(\lambda_\rho) = 
- g^{\mu\nu} + \frac{k^{\mu}k^{\nu}}{m_{\rho}^2} \,.
\en
In this process the contribution of neutral vector mesons with $I^J (J^{PC}) = 0^-(1^{--})$ (like $\omega^0$ 
meson) are strongly suppressed due to the $G$-parity conservation. 

\subsection{The sub-dominant process}

Second process is a reaction with exchange 
of $\rho^+$ meson or loop process with $\pi^0\pi^+$ exchange. This 
process is suppressed if one compares it with the first process which was considered before in Sec.~\ref{SecDominantCh}. This difference can be explained due to the typical  factors arising  from propagators 
$1/(t-m_\pi^2)^2$ and $1/(t-m_\rho^2)^2$,  at small negative $t$. In particular, the suppression factor  is proportional to the  $\sim m_\pi^4/m_\rho^4\sim 10^{-3}$ term. 
The matrix element  squared  of the process with  $\rho^+$ meson  exchange is
\eq
\frac{1}{2} \sum_{s(P_1)}\sum_{\lambda'} |M|^2 &&=  \frac{g_{\rho NN}^2 g_{\rho\rho\pi}^2}{2(t-m_\rho^2)^2}\Big[ 
2 m_N^2 (m_\pi^2 - m_\rho^2)^2  \nn \\
&& + 
 2 E_{k1} m_N (m_\pi^2- m_\rho^2)^2 (x-1) \nn \\
&&
+  E_{k1} (m_\pi^2 - m_\rho^2) (1 + x)  \\
&& - 
 4 E_{k1}^2 m_N^2 (m_\pi^2 + m_\rho^2) (x-1)^2 \nn \\
&&
+ 
 2 m_N t (-2 m_N (m_\pi^2 + m_\rho^2) 
 \nn \\
&&
+ 2 m_N^2 t^2    + 2 E_{k1}^2 m_N (1 + x^2))
\Big] \,. \nn
\en
 This channel  provides a  negligible yield of $\rho^0$  meson, so that  we do not take into  account this term for the calculation of the the  bounds on  dark photon  parameter space. 
\vspace{0.5cm}

\section{Bounds}
\label{bounds_sec}

Cosmology argument connected to nature of thermal
DM in the early Universe enclosed in relation between the
measured DM relic density and model parameters. In order to illustrate the results on the expected reach of NA64$_h$ we introduce the  dimensionless parameter 
$y=\alpha_D\epsilon^2(m_\chi/m_{A'})^4$ \cite{Berlin:2018bsc} which is convenient to use for the thermal target 
DM parameter space.  In particular,  by exploiting this parameter  $y=\alpha_D\epsilon^2(m_\chi/m_{A'})^4$, we 
can compare  the existed and projected  limits of NA64$_h$  with the typical relic DM parameter space. 

In Fig.~\ref{DiagDiffAlphaD}  we show  the constraints at $90\%$~CL on dark photon couplings from neutral vector meson for conservative number of  pions on target $10^9$ (few days of data taking)
and projected future statistics that corresponds
 to the  $5\times 10^{12}$ pions on target implying the NA64$_h$ pion 
beam  design.
Limits are derived for two benchmark sets
 of the DM parameters: 
$\alpha_D=0.5$ and $\alpha_D=0.1$ for $m_{A'}=3m_\chi$.
The typical limit for the  conservative  statistics is comparable with bounds 
which  are obtained from direct the production of $\omega$ and $\phi$ mesons at $e^+ e^-$ 
collider  \cite{BESIII:2018bec}.
The same results have been also obtained for the case of electron fixed target experiments 
(NA64$_e$ and LDMX) that implies  the  search for DM in the  missing energy signatures described in Ref.~\cite{Schuster:2021mlr}. 

Moreover, for the dark photons it is important to obtain the bound on  the  kinetic mixing parameter $\epsilon$ that is  originated from the mixing of hidden spin-1 boson with $\rho^0$ meson. That type of coupling results in the invisible vector meson decay  to dark photon 
$\rho \to A' \to \chi \bar{\chi}$.  	
Here we would like to note that for projected statistics  $\sim 5 \times 10^{12}$ $\pi\mbox{OT}$ the direct dark photon production results 
in a  relatively weak bounds of the kinetic mixing parameter as we have from NA64$_e$ at current statistics. The resonant production of DM is more effective  due to the amplified  magnitude of the cross-section  near the resonant dark photon  mass term. 
In this area bounds are more strict. 

\begin{figure}[t]
	\includegraphics[width=0.49\textwidth,clip]{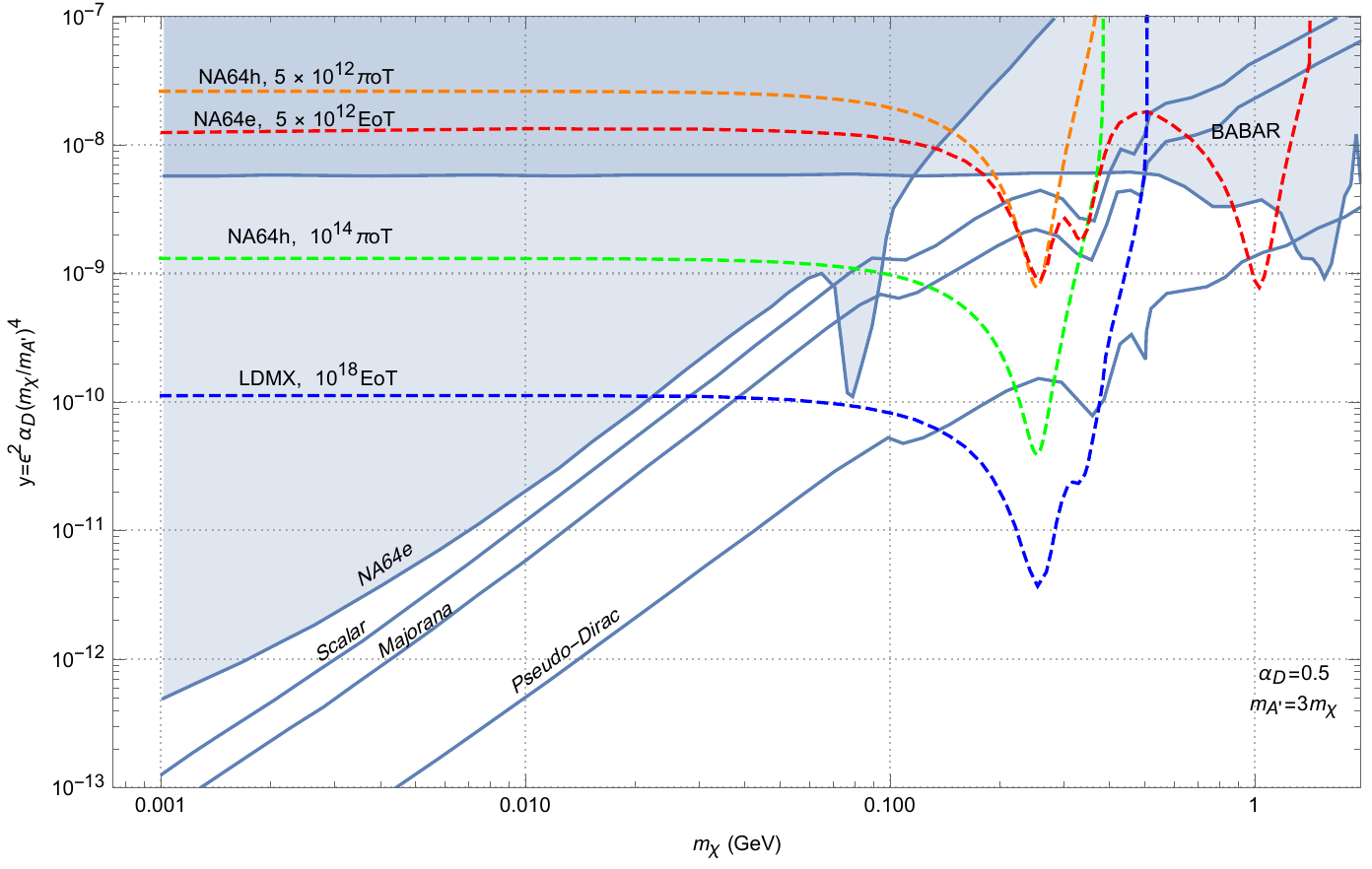}
 	\includegraphics[width=0.49\textwidth,clip]{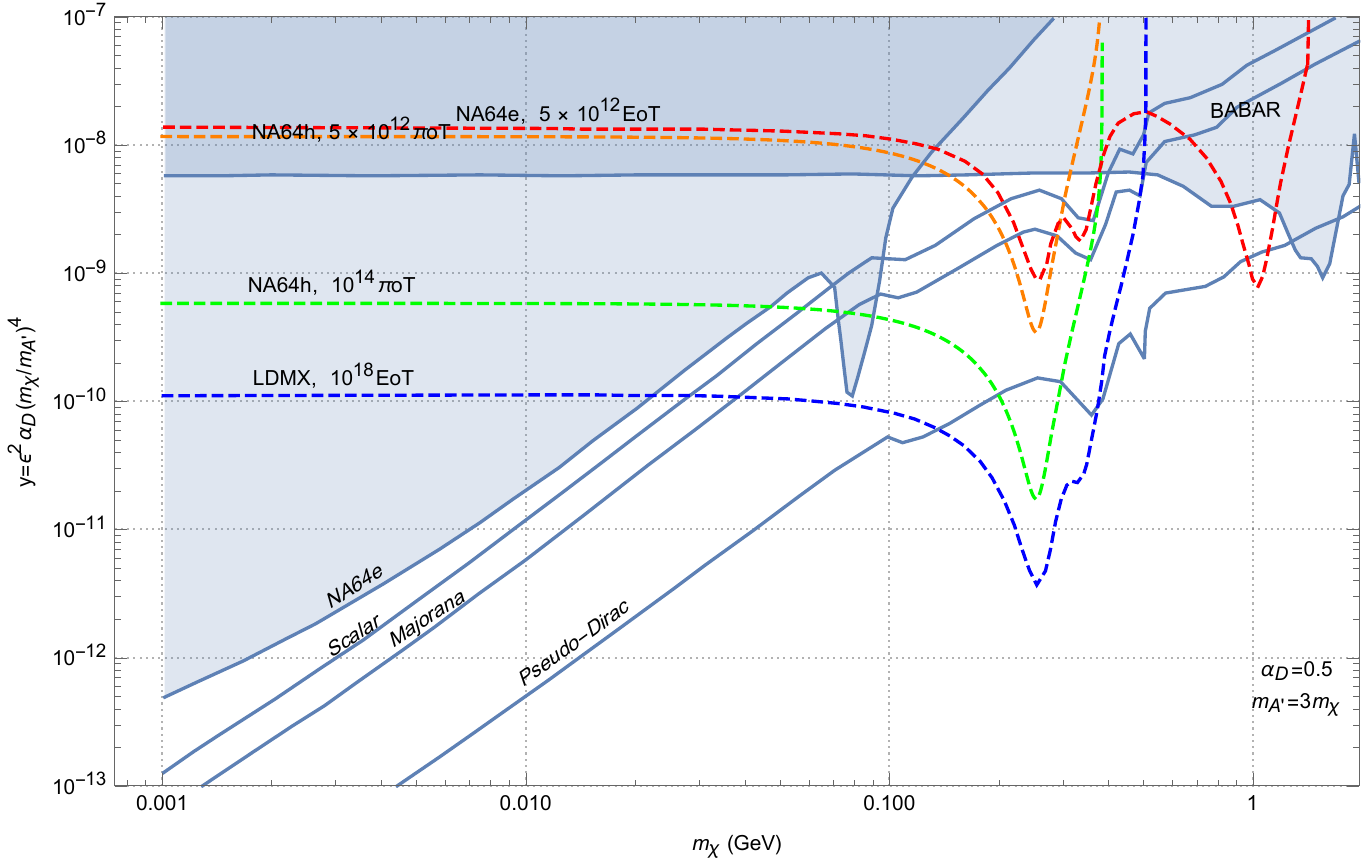}
	\caption{ The NA64$_h$ projected constraints for $5\times 10^{12}$ and $10^{14}$ pion on target and the expected reach of NA64$_e$ and LDMX experiments  with electron beam \cite{Schuster:2021mlr}   for $\alpha_D=0.5$ and $m_{A'}=3m_\chi$. In left panel we imply for NA64$_h$ that the recoil energy transferred from pion to the nucleon can be as small as  0.8  GeV. In right panel show the case of 1.2 GeV for  the recoil energy of nucleon.}
	\label{Ultimate}
\end{figure}

We should underline important advantages for analyzing invisible vector meson decay by using pion beams and missing energy techniques in the fixed target experiments. In particular, for $\rho^0$ meson production the  $\pi^+\pi^-$ channel 
is dominant~\cite{HADES:2020kce,Manley:1984jz}. Besides, this invisible decay of $\rho^0$ meson coupled with dark photon implies the interaction coupling of spin-1 hidden boson with quarks. 
As a result, for the ultimate statistics of NA64$_h$ at the level of  $\simeq 10^{14}\, \pi\mbox{OT}$
one can obtain a relatively strong bounds on the typical DM thermal target parameter $y=\alpha_D\epsilon^2(m_\chi/m_{A'})^4$, that can be better than the expected ultimate limit for electron beam of the NA64$_e$ experiment. 
For statistics of $\sim 5 \times 10^{12} \pi$OT bounds will be similar 
which were obtained for ultimate statistics of NA64$_e$ experiment. Note, that optimistic bounds from future statistics for pion 
beam can be comparable with the expected reach from invisible meson decay for ultimate statistics of NA64$_e$. Nevertheless, the optimistic bound of LDMX ($10^{18}$ EOT) can rule out the expected reach of NA64$_h$ for the ultimate statistics at the level of $10^{14}\, \pi\mbox{OT}$  (see, e.~g.~Ref.~\cite{Schuster:2021mlr} for detail).
The comparisons of the regarding expected limits are shown in Fig.~\ref{Ultimate}. 
In this picture we use the bounds obtained in 
Ref.~\cite{Schuster:2021mlr} for the ultimate limit of invisible vector decay at NA64$_e$ and LDMX experiments. 
These bounds for the LDMX experiment include limits from $\rho$ and $\omega$ meson, for the NA64 experiment bound includes 
limits from $J/\psi$ invisible decay too. Besides, there needs to note a difference in couplings between vector meson and dark 
photon with \cite{Schuster:2021mlr}. We exploit  the  effective field theory to fix meson couplings. 
We plan to expand our analysis of vector meson 
production for pion beam scattering at fixed target including numerical simulation and analysis of addition vector mesons. 
Besides, we also plan  to consider real pQCD calculation in our analysis of fix target experiments with pion beam energy  
at 50 GeV or 100 GeV. 
\onecolumngrid
\onecolumngrid
\twocolumngrid
\twocolumngrid
\quad
\\
\newline

In the framework of the proposal, we note that missing energy techniques require an  approximate zero background for 
identification of missing energy signals. From Table.~\ref{ParamTable}, one can see that decreasing 
beam energy to 20 GeV can increase the expected limit from invisible meson decay at NA64$_h$ experiment.  

\begin{figure}[t]
	\includegraphics[width=0.49\textwidth,clip]{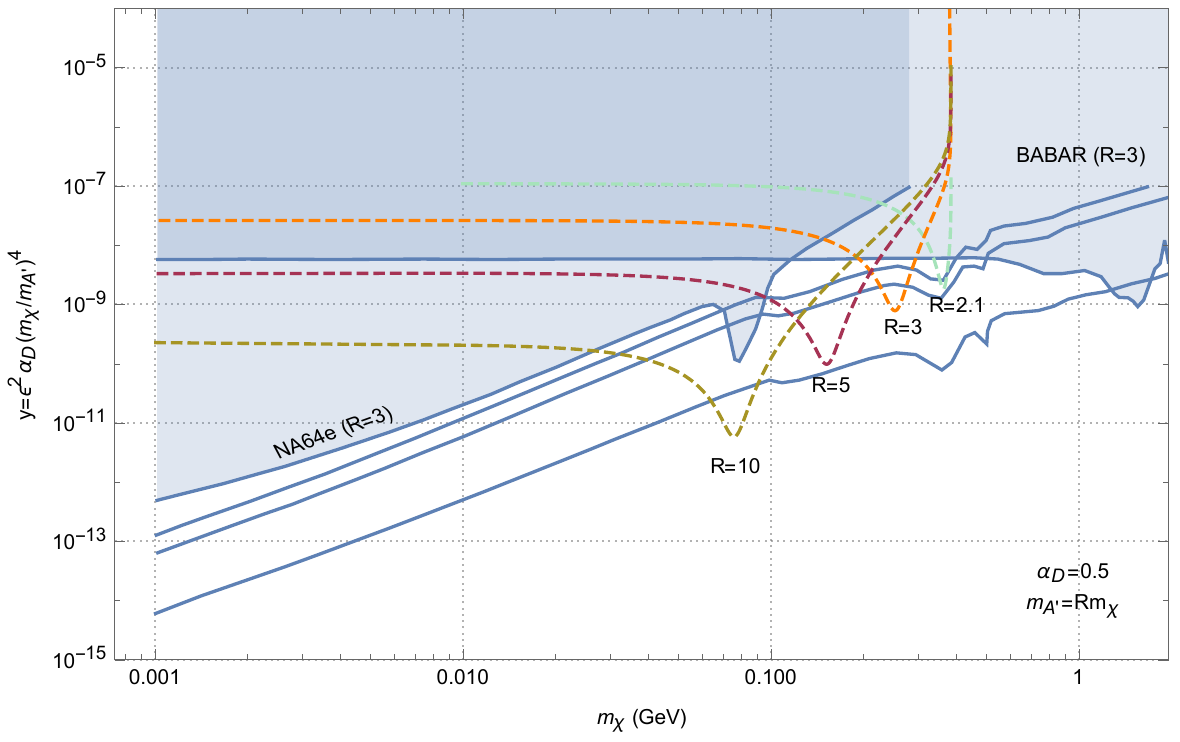}
	\caption{Projected 90\% C.L. exclusion for statistics $5 \times 10^{12}$ pions on target from invisible $\rho^0$ neutral vector meson decay into pseudo-Dirac DM by transition via dark photon. Constraint is presented for several choices of mass ratio $R=m_{A'}/m_\chi$. Thermal targets and experimental bounds are shown for $R=3$.}
	\label{DiagdiffR}
\end{figure}

In Fig.~\ref{DiagdiffR} we show  the sensitivity curves for various mass ratio $R=m_{A'}/m_\chi$ for the  dark photons and 
pseudo-Dirac fermions implying  the projected ultimate statistics $\sim 5\times 10^{12}$ of pions on target.  
 Our results are in full agreement with one presented in Ref.~\cite{Schuster:2021mlr}. In particular, the larger value of the 
 $R$,   the smaller typical resonant masses of DM.  Moreover,  one can achieve the better limits for large value of $R$ 
 parameter,  that follows from the Breit-Wigner shape for resonance production. 
 
 In addition to everything mentioned before, one can obtain the typical bound on the invisible branching 
 \eq
 \mathrm{Br}(\rho^0\to \text{inv.}) &<& 1.2 \times 10^{-3} \quad \textnormal{for $ 3.3 \times 10^{9}$ $ \pi$OT} \quad\\
 \mathrm{Br}(\rho^0\to \text{inv.}) &<& 7.5 \times 10^{-7} \quad \textnormal{for $5\times 10^{12} $ $\pi$OT} \quad
 \en
 in the framework 90\% C.L. of missing energy signature implying zero signal events and  background free case. 
 
\section{Conclusion}	
\label{Conclusion}

Invisible decay of the $\rho^0$ meson was studied by using the missing energy design of the fixed-target experiment with pion beam. We used the NA64 hadronic design with $\pi^-$ beam scattered in hadronic calorimeter that serves as a target. 
We derived the bounds on parameter space of pseudo-Dirac DM for the proposed conservative and ultimate statistics of pions on target. We compared the regarding expected reach 
with typical curves of DM relic density. 
We analyzed two possibilities of the pion beam energy: 50 GeV and 100 GeV. 
The second possibility of the pion beam energy (100 GeV) requires a more narrow angle
of meson production if we want to search for missing energy signals at small recoil energy in the background. 
All these cuts lead to less meson yield at high energy of pion beam. 
Wherein, we note that decreasing the pion energy beam can give a chance to obtain 
a more strict limit to parameter DM using invisible mode of vector vector meson. 
This was tested numerically for the specific value of the pion beam energy equal to 
20 GeV. Additionally we would like to note that we propose to use a potential of missing energy techniques in the fixed target experiment at PS/CERN with pion beam energy 
of 6 GeV. Such a configuration should be more optimal for 
study considered charge-exchange processes with vector meson production.    

Obtained results in the present paper led to optimistic bounds which can be made 
by analysis of invisible vector meson decay as a signal to possible DM production. 
We showed that the pion beam can be an effective tool for study  of DM by missing energy/momenta technique in the fixed target experiments. 
In future we plan to make a more comprehensive analysis of detector in the 
setup of the fixed target experiment with pion beam. 
Besides, we plan to study meson production at high energies in fixed-target experiments with pion beams by using both the model-independent and model-independent techniques. 
	
\begin{acknowledgments} 
We would like to thank S. Ershov for discussion. The work of A.~S.~Zh. on exclusion limits calculation for the fixed target experiments is supported by Russian Science Foundation
(grant No. RSF 23-22-00041).  The work of A.~S.~Zh. under Sec. II and III is  supported by the Foundation for the Advancement of Theoretical Physics and Mathematics
"BASIS". The work of D.~V.~Kirpichnikov on calculation of signal missing energy events of NA64$_h$ is supported by  Russian Science Foundation (grant No. RSF 21-12-0037).
The work was funded by ANID PIA/APOYO 
AFB220004 (Chile), by FONDECYT (Chile) under Grant No. 1230160, 
and by ANID$-$Millen\-nium Program$-$ICN2019\_044 (Chile).
\end{acknowledgments}

\bibliography{PipiModeBib.bib}

\end{document}